\documentclass[conference]{IEEEtran}

\usepackage[T1]{fontenc}
\usepackage[utf8]{inputenc}
\usepackage{amssymb}
\usepackage[pdftex]{graphicx}
\usepackage{subcaption} % for subfigures
\usepackage{amsmath}
\usepackage{amsthm}
\usepackage{mathtools}
\usepackage[noadjust]{cite}
\usepackage[hidelinks]{hyperref}
\usepackage[acronym]{glossaries}
\usepackage{siunitx}
\usepackage{tikz}
\usepackage{flushend}
\usepackage{multirow}
\usepackage{pgfplots}
\usepackage[ruled,boxed,lined,linesnumbered]{algorithm2e}
\pgfplotsset{compat=1.17}

\usetikzlibrary{arrows,chains,matrix,positioning,scopes}
\usetikzlibrary{calc}
\usetikzlibrary{fit,shapes}

\newacronym{NPRACH}{NPRACH}{narrowband physical random-access channel}
\newacronym{ToA}{ToA}{time of arrival}
\newacronym{CFO}{CFO}{carrier frequency offset}
\newacronym{NBIoT}{NB-IoT}{narrowband internet of things}
\newacronym{5GNR}{5G NR}{5G New Radio}
\newacronym{3GPP}{3GPP}{3rd Generation Partnership Project}
\newacronym{UMi}{UMi}{urban microcell}
\newacronym{RMSE}{RMSE}{root-mean-square error}
\newacronym{NN}{NN}{neural network}
\newacronym{BS}{BS}{base station}
\newacronym{UE}{UE}{user equipment}
\newacronym{SG}{SG}{symbol group}
\newacronym{CP}{CP}{cyclic prefix}
\newacronym{OFDM}{OFDM}{orthogonal frequency division multiplexing}
\newacronym{FFT}{FFT}{fast Fourier transform}
\newacronym{AWGN}{AWGN}{additive white Gaussian noise}
\newacronym{DFT}{DFT}{discrete Fourier transform}
\newacronym{FNR}{FNR}{false negative rate}
\newacronym{FPR}{FPR}{false positive rate}
\newacronym{RG}{RG}{resource grid}
\newacronym{RE}{RE}{resource element}
\newacronym{SNR}{SNR}{signal-to-noise ratio}
\newacronym{1D}{1D}{one-dimensional}
\newacronym{MLP}{MLP}{multilayer perceptron}
\newacronym{BCE}{BCE}{binary cross-entropy}
\newacronym{KL}{KL}{Kullback–Leibler}
\newacronym{SGD}{SGD}{stochastic gradient descent}
\newacronym{ppm}{ppm}{parts-per-million}
\newacronym{ICI}{ICI}{inter-carrier interference}
\newacronym{GNN}{GNN}{graph neural network}
\newacronym{BP}{BP}{belief propagation}
\newacronym{FEC}{FEC}{forward error correction}
\newacronym{ML}{ML}{maximum likelihood}
\newacronym{LDPC}{LDPC}{low-density parity-check}
\newacronym{HDPC}{HDPC}{high-density parity-check}
\newacronym{SCL}{SCL}{successive cancellation list}
\newacronym{SC}{SC}{successive cancellation}
\newacronym{URLLC}{URLLC}{ultra-reliable low-latency communications}
\newacronym{APP}{APP}{a posterior probability}
\newacronym{MIMO}{MIMO}{multiple-input multiple-output}
\newacronym{CNN}{CNN}{convolutional neural network}
\newacronym{BER}{BER}{bit error rate}
\newacronym{BPSK}{BPSK}{binary phase shift keying}
\newacronym{LLR}{LLR}{log-likelihood ratio}
\newacronym{FN}{FN}{factor node}
\newacronym{VN}{VN}{variable node}
\newacronym{CN}{CN}{check node}
\newacronym{MPNN}{MPNN}{message passing neural network}

\renewcommand{\vec}[1]{\mathbf{#1}}
\newcommand{\vecs}[1]{\boldsymbol{#1}}

\newcommand{\cv}{\vec{c}}

\newcommand{\gv}{\vec{g}}
\newcommand{\hv}{\vec{h}}

\newcommand{\mv}{\vec{m}}
\newcommand{\nv}{\vec{n}}

\newcommand{\qv}{\vec{q}}

\newcommand{\wv}{\vec{w}}
\newcommand{\xv}{\vec{x}}
\newcommand{\yv}{\vec{y}}

\newcommand{\thetav}{\vecs{\theta}}

\newcommand{\ellv}{\vecs{\ell}}

\newcommand{\Hm}{\vec{H}}

\definecolor{mittelblau}{RGB}{0, 126, 198}
\definecolor{violettblau}{cmyk}{0.9, 0.6, 0, 0}
\definecolor{rot}{RGB}{238, 28 35}
\definecolor{apfelgruen}{RGB}{140, 198, 62}
\definecolor{gelb}{RGB}{1, 221, 0}
\definecolor{orange}{RGB}{244, 111, 33}
\definecolor{pink}{RGB}{237, 0, 140}
\definecolor{lila}{RGB}{128, 10, 145}
\definecolor{hellgrau}{RGB}{224, 224, 224}
\definecolor{mittelgrau}{RGB}{128, 128, 128}
\definecolor{dunkelgrau}{RGB}{80,80,80}
\definecolor{anthrazit}{RGB}{19, 31, 31}

\pgfplotscreateplotcyclelist{corporate colours markers}{%
rot, every mark/.append style={fill=.!80!rot},mark=*\\%
mittelblau, every mark/.append style={fill=.!80!mittelblau},mark=square*\\%
apfelgruen, every mark/.append style={fill=.!80!apfelgruen},mark=triangle*\\%
orange, mark=star\\%
pink, every mark/.append style={fill=.!80!pink},mark=diamond*\\%
violettblau, every mark/.append style={fill=.!80!violettblau},mark=otimes*\\%
lila, mark=|\\%
gelb, every mark/.append style={fill=.!80!gelb},mark=pentagon*\\%
hellgrau, mark=text,text mark=p\\%
anthrazit, mark=text,text mark=a\\%
}

\pgfplotscreateplotcyclelist{corporate colours markers double}{%
rot, every mark/.append style={fill=.!80!rot},mark=*\\%
rot, dashed, every mark/.append style={fill=.!80!rot,solid},mark=*\\%
mittelblau, every mark/.append style={fill=.!80!mittelblau},mark=square*\\%
mittelblau, dashed, every mark/.append style={fill=.!80!mittelblau,solid},mark=square*\\%
apfelgruen, every mark/.append style={fill=.!80!apfelgruen},mark=triangle*\\%
apfelgruen, dashed, every mark/.append style={fill=.!80!apfelgruen,solid},mark=triangle*\\%
orange, mark=star\\%
orange, dashed, mark=star\\%
pink, every mark/.append style={fill=.!80!pink},mark=diamond*\\%
pink, dashed, every mark/.append style={fill=.!80!pink,solid},mark=diamond*\\%
violettblau, every mark/.append style={fill=.!80!violettblau},mark=otimes*\\%
violettblau, dashed, every mark/.append style={fill=.!80!violettblau,solid},mark=otimes*\\%
lila, mark=|\\%
lila, dashed, mark=|\\%
gelb, every mark/.append style={fill=.!80!gelb},mark=pentagon*\\%
gelb, dashed, every mark/.append style={fill=.!80!gelb,solid},mark=pentagon*\\%
}

\begin{document}

% \vspace*{2cm}
% Results:
% \begin{itemize}
% 	\item BCH\_10 best result without attributes
% 	\item BCH\_11 including small attributes
% 	\item BCH\_12 more attributes
% 	%\item good results for #load_weights(e2e_gnn, "log/BCH_19_50000.npy") + few k iters with 1-4
% 	%\item 7.2k weights (BCH_Ray_1657225718); maybe train with tanh? resimulated in BCH_18
% \end{itemize}

% \newpage

% Top matter
\title{Graph Neural Networks for Channel Decoding}

\author{\IEEEauthorblockN{Sebastian Cammerer, Jakob Hoydis, Fay\c{c}al A\"{i}t Aoudia, and Alexander Keller}
		\IEEEauthorblockA{NVIDIA\\Contact: scammerer@nvidia.com}
}

\maketitle

\begin{abstract}
In this work, we propose a fully differentiable \gls{GNN}-based architecture for channel decoding and showcase a competitive decoding performance for various coding schemes, such as \gls{LDPC} and BCH codes. The idea is to let a \gls{NN} learn a generalized message passing algorithm over a given graph that represents the \gls{FEC} code structure by replacing node and edge message updates with trainable functions.
Contrary to many other deep learning-based decoding approaches, the proposed solution enjoys scalability to arbitrary block lengths and the training is not limited by the curse of dimensionality.
We benchmark our proposed decoder against state-of-the-art in \emph{conventional} channel decoding as well as against recent deep learning-based results. For the (63,45) BCH code, our solution outperforms weighted \gls{BP} decoding by approximately 0.4\,dB with significantly less decoding iterations and even for 5G NR \gls{LDPC} codes, we observe a competitive performance when compared to conventional \gls{BP} decoding. For the BCH codes, the resulting \gls{GNN} decoder can be fully parametrized with only 9640 weights. %Finally, we make the source code of the proposed framework online available.
\end{abstract}
\glsresetall

\section{Introduction}

\emph{Codes on graphs} \cite{forney2001codes,richardson2008modern,kschischang2001factor} have become the workhorse of modern \gls{FEC} schemes and have, thereby, brought the theoretical asymptotic performance limits of channel coding within practical reach \cite{chung2001design,arikan2009channel,kudekar2013spatially}. As a result, many practical coding schemes in today's wireless communication systems can be interpreted as (sparse) graphs which inherently provide a low-complexity, iterative message passing decoder---namely, \gls{BP} decoding \cite{pearl1982reverend,gallager1962low}.
However, when focusing on short to intermediate codeword lengths, the decoding performance under \gls{BP} decoding shows a non-negligible gap when compared to \gls{ML} decoding for most codes \cite{liva2016code}.
For this regime, there is still no universal iterative coding scheme as either the decoding graph becomes dense (e.g., for Polar decoding \cite{cammerer2018sparse}) and/or the decoding graph has many short cycles.
For instance, finding an iterative decoding scheme for short length CRC-aided Polar codes is an active field of research \cite{geiselhart2021automorphism}.
Further, finding universal decoding algorithms for Polar, \gls{LDPC}, and BCH codes in the same hardware engine may simplify the chip design of future wireless systems.

In the recent years, the field of deep learning for communications has rapidly progressed and showcased several impressive results ranging from fully \gls{NN}-based \gls{OFDM} receivers \cite{honkala2021deeprx} via \emph{weighted} \gls{BP} decoding \cite{nachmani2016learning} up to end-to-end optimization of the complete transceiver signal processing chain \cite{o2017introduction}. However, even though the idea of incorporating expert domain knowledge in the \gls{NN} structure has been investigated many times \cite{nachmani2016learning,cammerer2020trainable}, there is still no structured way of embedding a known graph (or data) structure in current deep learning architectures.
Deep learning for communications is therefore often doomed by the \emph{curse of dimensionality} \cite{wang1996artificial,gruber2017deep}, meaning that the training complexity is exponential in the number of information bits. As a result, most current learning approaches in the literature \cite{bennatan2018deep, choukroun2022error} besides weighted BP decoding \cite{nachmani2016learning} are limited by their scalability and tend to have a tremendous amount of trainable parameters (in the range of millions of weights even for short codes). These schemes require expensive re-training--or to store and load pre-calculated parameters--for every set of new code parameters.

From a more philosophical point-of-view, we have not yet found the \gls{NN} architectures that suit well to channel coding tasks, similar to what \glspl{CNN} are for computer vision and transformers are for natural language processing, respectively.
Thus, we believe that the obvious close relations between factor graphs \cite{kschischang2001factor} and \glspl{GNN} \cite{satorras2021neural}, render them into a promising architecture for trainable decoders. We refer the interested reader to \cite{cappart2021combinatorial} for a more general view on \glspl{GNN}.

The idea of using \glspl{MPNN} \cite{gilmer2017mpnn} for channel decoding has been recently introduced in \cite{satorras2021neural} and the authors of \cite{liao2022gnn} use \glspl{GNN} to support the design of Polar codes. Inspired by these results, we aim to provide an introduction to \glspl{GNN} for channel decoding tasks and to understand their advantages and limitations. Please note that the authors in \cite{satorras2021neural} \emph{augment} \gls{BP} decoding by the combination with a \gls{GNN} while in our work, we propose a fully learned decoder.
Our main contribution is to
\begin{itemize}
	\item Present a fully differentiable \gls{GNN}-based decoder for linear binary block codes;
 	\item Clarify terminology and \emph{translate} the \gls{GNN} terminology to the channel coding domain;
 	\item Benchmark against state-of-the-art 5G NR \gls{LDPC} and BCH codes;
  	\item Provide design guidelines for \gls{GNN} decoders and interpretation of an extensive hyperparameter optimization;
  	\item Support reproducible research: the source code is implemented with Sionna \cite{hoydis2022sionna} and is available online.\footnote{\url{https://github.com/NVlabs/gnn-decoder}.}
\end{itemize}

It is worth a note that many of today's problems in communications can be formulated by the framework of \emph{factor graphs} \cite{kschischang2001factor, schmid2022low}. We would like to emphasize our fascination for this omnipresence of graphs in many communications problems and the fact that we still do not have a universal framework to embed such graphs in our \gls{NN} designs. We intend this paper to be a first step towards such a framework. However, we do not want to oversell the results in this well-investigated field of channel coding which has already seen many powerful hand-crafted solutions. We rather hope this paper opens more research questions than it answers.

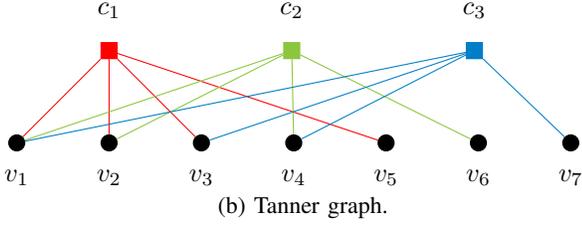
\begin{figure}
	\centering
	\begin{subfigure}[t]{0.3\textwidth}
		$$\Hm=
		\begin{bmatrix}
		\color{red}1 & \color{red} 1 & \color{red} 1 & 0 & \color{red} 1 & 0 & 0  \\
		\color{apfelgruen} 1 & \color{apfelgruen} 1 & 0 & \color{apfelgruen} 1 & 0 & \color{apfelgruen} 1 & 0  \\
		\color{mittelblau} 1 & 0 &  \color{mittelblau} 1 & \color{mittelblau} 1 & 0 & 0 & \color{mittelblau} 1
		\color{black}
		\end{bmatrix}
		$$
		\vspace*{-0.1cm}
		\caption{Parity-check matrix.}
	\end{subfigure}
	%\vspace*{0.4cm}
	\begin{subfigure}[t]{\columnwidth}
		\centering
		%TIKZ AWGN-Channel image
\tikzstyle{cnd} = [rectangle,draw=black,fill=black,thick,text width=2mm,text height=2mm,inner sep=0pt]
\tikzstyle{vnd} = [circle,draw=black,fill=black,thick,text width=2mm,inner sep=0pt]

\begin{tikzpicture}[scale=0.6]

		%VND
	    \node [vnd, label={[yshift=-0.8cm,xshift=-0.cm]$v_1$}] (vnd1) {};
	    \node [vnd, right=1cm of vnd1, label={[yshift=-0.8cm,xshift=-0.cm]$v_2$}] (vnd2) {};
	    \node [vnd, right=1cm of vnd2, label={[yshift=-0.8cm,xshift=-0.cm]$v_3$}] (vnd3) {};
	    \node [vnd, right=1cm of vnd3, label={[yshift=-0.8cm,xshift=-0.cm]$v_4$}] (vnd4) {};
	    \node [vnd, right=1cm of vnd4, label={[yshift=-0.8cm,xshift=-0.cm]$v_5$}] (vnd5) {};
	    \node [vnd, right=1cm of vnd5, label={[yshift=-0.8cm,xshift=-0.cm]$v_6$}] (vnd6) {};
	    \node [vnd, right=1cm of vnd6, label={[yshift=-0.8cm,xshift=-0.cm]$v_7$}] (vnd7) {};

	    %CND
	    \node [cnd,red, above=1cm of vnd2, label={[yshift=0.2cm,xshift=0.cm]$c_1$}] (cnd1) {};
	    \node [cnd,apfelgruen, right=2.2cm of cnd1, label={[yshift=0.2cm,xshift=0.cm]$c_2$}] (cnd2) {};
	    \node [cnd, mittelblau,right=2.2cm of cnd2, label={[yshift=0.2cm,xshift=0.cm]$c_3$}] (cnd3) {};

		%startpoint
		%\node [cnd, fill=white, draw=none, left=1cm of vnd1, inner sep=0pt] (start) {};

	    %draw edges
	    %CND1
	   \path[-,red] (vnd1) edge node{} (cnd1);
	   \path[-,red] (vnd2) edge node{} (cnd1);
	   \path[-,red] (vnd3) edge node{} (cnd1);
	   \path[-,red] (vnd5) edge node{} (cnd1);

	  %CND2
	  \path[-,apfelgruen] (vnd1) edge node{} (cnd2);
	  \path[-,apfelgruen] (vnd2) edge node{} (cnd2);
	  \path[-,apfelgruen] (vnd4) edge node{} (cnd2);
	  \path[-,apfelgruen] (vnd6) edge node{} (cnd2);

	  %CND3
	  \path[-,mittelblau] (vnd1) edge node{} (cnd3);
	  \path[-,mittelblau] (vnd3) edge node{} (cnd3);
	  \path[-,mittelblau] (vnd4) edge node{} (cnd3);
	  \path[-,mittelblau] (vnd7) edge node{} (cnd3);

	    %text
	 %\node[draw=none,below=0.7cm of vnd1] (vndtext){Variable nodes (VN)};
	 %\node[draw=none,above=1.9cm of vnd1](cndtext){Check nodes (CN)};

\end{tikzpicture}
		\vspace*{-0.2cm}
		\caption{Tanner graph.}
	\end{subfigure}
	\caption{Bipartite graph for the (7,4) Hamming code.}
	\label{fig:bipartite_hamming}
	\vspace*{-0.2cm}
\end{figure}

\section{GNNs for Channel Decoding}

We consider a binary linear block code $\mathcal{C}$ of length $n$ that can be described by a binary parity-check matrix $\Hm^{(n-k) \times n} \in \mathbb{F}_2^{(n-k)\times n}$ where each codeword represents $k$ information bits.\footnote{For simplicity, we assume $\Hm$ is full rank.}
A binary vector is a codeword $\cv \in \mathbb{F}_2^n$ of $\mathcal{C}$ \emph{iff} in GF(2) it holds that $\Hm \cv^\text{T} = \mathbf{0}$.
For the \gls{AWGN} channel\footnote{Extensions to other channels are straightforward.} with noise variance $\sigma_\text{ch}^2$ and \gls{BPSK} modulation $\xv = -2\cv+1$, the receiver observes $\yv = \xv + \nv$ with $\nv \sim \mathcal{N}(\mathbf{0}, \sigma_\text{ch}^2 \mathbf{I})$. After demapping, the \gls{LLR} of the $i$th codeword bit $c_i$ is given as
$$\ell_{\text{ch},i} = \operatorname{log}\frac{p(c_i=0)}{p(c_i=1)} = \frac{2y_i}{\sigma_\text{ch}^2}$$
and $\ellv_\text{ch}$ denotes the vector of all $n$ channel \glspl{LLR}.

The optimal bitwise \gls{ML} decoder performs
$$\hat{c_i} = \operatorname{argmax}_{c_i \in \{0,1\}} p(\yv |c_i) \quad \forall i =0,\dots,n-1.$$
It is worth mentioning, that some decoders provide \gls{ML}-\emph{sequence} estimation (i.e., minimize the BLER) such as the Viterbi decoder while others  are \gls{ML} \emph{bitwise} estimators (i.e., minimize the BER) such as the BCJR decoder. As a result of bitwise \gls{ML} decoding, $\hat{\cv}$ is not necessarily a valid codeword.

By interpreting $\Hm$ as bipartite graph \cite{tanner1981recursive}, we immediately get a decoding graph for message passing decoding. As illustrated in Fig.~\ref{fig:bipartite_hamming}, each row of $\Hm$ represents a single parity-check (so-called constraint or \gls{CN}) and each column represents a codeword position (so-called \gls{VN}). For finite length, \gls{BP} is known to be suboptimal due to cycles in the decoding graph (no tree structure).
For a detailed introduction to \gls{BP} decoding, we refer to \cite{ryan2004introduction,richardson2008modern} and would like to emphasize a few important properties:
\begin{itemize}
	\item $\Hm$ is not unique and can be modified (e.g., by the linear combination of rows), i.e., multiple graphs can be assigned to the same code (see \cite{forney2001codes}). Thus, several graph properties such as specific cycles or node degrees are strictly speaking \emph{not} code properties.
 	\item $\Hm$ can be also \emph{overcomplete}, i.e., more than $n-k$ rows can be part of $\Hm$ which are then linearly dependent (see \cite{buchberger2020pruning} for a practical application in decoding).
 	\item The set of all $2^k$ codewords $\cv$ defines the code $\mathcal{C}$. However, for a linear code, the code is fully defined by $k$ linearly independent codewords (see curse of dimensionality in \cite{gruber2017deep}).
  	\item For a linear code, the all-zero codeword $\cv=\mathbf{0}$ is always part of the code as $\Hm\mathbf{0}^\text{T}=\mathbf{0}$.
\end{itemize}

\subsection{Bipartite Graph Neural Network Framework}

\begin{figure*}
	\centering
	\vspace*{0.25cm}
	\resizebox{0.85\textwidth}{!}{\includegraphics[trim= 2cm 5.2cm 2cm 4cm]{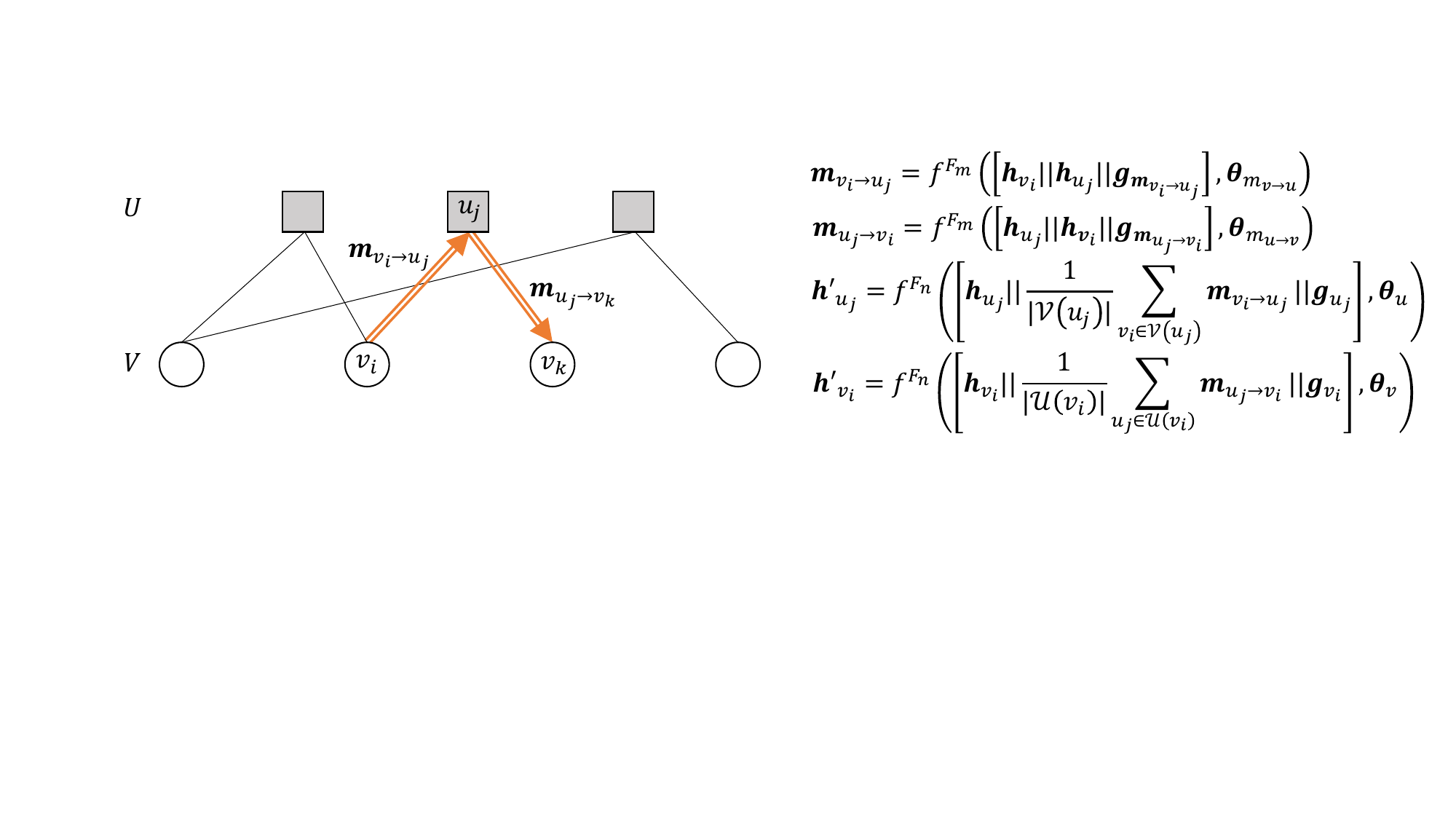}}
	\vspace*{-1.6cm}
	\caption{Graph neural network based message passing for bipartite graphs (see Alg.~\ref{alg:gnn_dec}).}
	\label{fig:mp_bp_graph}
	%\vspace*{-0.2cm}
\end{figure*}

We now describe the bipartite \gls{GNN}-based message passing decoding algorithm for a given parity-check matrix $\Hm$ following \cite{gilmer2017mpnn}. %\footnote{To support reproducible research, the source code is implemented with Sionna \cite{hoydis2022sionna} and online available as \textcolor{red}{*tbd*}.}
Learning the graph structure itself is also an active field of research, but not considered in this work.

Let $f^F(\xv,\thetav_\text{NN})$ denote a parametrized function that maps $f^F: \xv \mapsto \mathbb{R}^F$ and can be parametrized by the weights $\thetav_\text{NN}$ (e.g., we use simple \glspl{MLP} throughout this work and other options are possible). Note that the superscript $F \in \mathbb{N}$ defines the dimensionality of the messages.
As illustrated in Fig.~\ref{fig:mp_bp_graph}, we operate on a bipartite graph and we introduce the following nodes/functions
\begin{itemize}
	\item \textbf{\Glsfirst{VN}:} Each node $v_i$ represents a transmitted codeword bit $c_i$. Let $\mathcal{V}$ denote the set of all \glspl{VN}.
 	\item \textbf{\Gls{FN}:} Each node $u_j$ represents one of the $n-k$ code constraints (i.e., check nodes in \gls{BP} decoding). Let $\mathcal{U}$ denote the set of all \glspl{FN}.
  	\item The \gls{VN} $v_i$ is connected to the \gls{FN} $u_j$ (and vice versa) \emph{iff} $H_{j,i}=1$.
    \item $\mathcal{V}(u_j)$ denotes the neighborhood of $u_j$, i.e., the set of all variable nodes connected to factor node $u_j$.
    \item $\mathcal{U}(v_i)$ denotes the neighborhood of $v_i$, i.e., the set of all factor nodes connected to variable node $v_i$.
    \item $\mathcal{E}$ denotes the set of all edges $e_{i,j} = (v_i, u_j)$ in the graph, i.e., $ e_{i,j} = (v_i, u_j) \in \mathcal{E} \Leftrightarrow H_{j,i}=1$.
    \item \textbf{Node values:} Each node computes an $F_\text{n}$-dimensional vector $\hv_{v_i,}\hv_{u_j}\in \mathbb{R}^{F_\text{n}}$ for each \gls{VN} $v_i$ and \gls{FN} $u_j$, respectively.
    \item \textbf{(Directed) edge values:} Each directed edge message has an assigned $F_\text{m}$-dimensional value $\mv_{v_i \to u_j},\mv_{u_i \to v_j} \in \mathbb{R}^{F_\text{m}}$. Remark: the values are (generally) different for each direction.
    \item \textbf{Node attributes:} Every node can have an assigned node attribute $\gv_{v_i}, \gv_{u_j} \in \mathbb{R}^{F_\text{na}}$ for \gls{VN} and \gls{FN}, respectively. This is node specific, i.e., in total $|\mathcal{V}|+|\mathcal{U}|=2n-k$ node attributes exist.
	\item \textbf{Edge attributes:} Every edge can have an assigned edge attribute $\gv_{m_{v_i \to u_j}}, \gv_{m_{u_j \to v_i}} \in \mathbb{R}^{F_\text{ma}}$, respectively. This is edge specific and depends on the edge direction, i.e., in total $2|\mathcal{E}|$ edge attributes exist. Remark: this is similar to the trainable weights in \cite{nachmani2016learning}.
\end{itemize}

We implement the \gls{GNN} decoding algorithm for bipartite graphs as \gls{MPNN} \cite{gilmer2017mpnn,satorras2021neural}. Therefore, we introduce two kind of trainable functions: (a) the \emph{edge message update functions} and (b) the \emph{node update functions}.
An updated message value from \gls{VN} $v_i$ to \gls{FN} $u_j$ can be calculated as
\begin{equation}
	\mv_{v_i \to u_j} = f^{F_\text{m}}([\hv_{v_i} || \hv_{u_j} || \gv_{m_{v_i \to u_j}}], \thetav_{m_{v \to u}})
	\label{eq:msg_vu_up}
\end{equation}
where $\thetav_{m_{v \to u}}$ denotes the trainable parameters of the node update and $[\xv||\yv]$ denotes the concatenation of the vectors $\xv$ and $\yv$. Please note that the same weights $\thetav_{m_{v \to u}}$ are applied to all edges.
Similarly, the updated message value from \gls{FN} $u_j$ to \gls{VN} $v_i$ is given as
\begin{equation}
	\mv_{u_j \to v_i} = f^{F_\text{n}}([\hv_{u_j} || \hv_{v_i} || \gv_{m_{u_j \to v_i}}], \thetav_{m_{u \to v}})
	\label{eq:msg_uv_up}
\end{equation}
with $\thetav_{m_{u \to v}}$ denoting another (shared) set of trainable parameters.
The updated node value for \gls{VN} $v_i$ is calculated as
\begin{equation}
	\hv'_{v_i} = f^{F_n}([\hv_{v_i}||\bigoplus_{u_j \in \mathcal{U}(v_i)} \mv_{u_j \to v_i}  ||\gv_{v_i}], \thetav_{v})
	\label{eq:node_v_up}
\end{equation}
with $\thetav_{v}$ being the (shared) parameters of the \gls{VN} update function and $\bigoplus$ denotes a message \emph{aggregation} function. Throughout this work, we use the \emph{sum} or the \emph{mean} operator for $\bigoplus$.\footnote{The aggregation function must be permutation invariant, i.e., invariant to the order in which incoming messages are processed. For simplicity, we assume \emph{mean} aggregation in the description of Alg.~\ref{alg:gnn_dec}}
Similarly, the \gls{FN} value update can be written as
\begin{equation}
	\hv'_{u_j} = f^{F_n}([\hv_{u_j}||\bigoplus_{v_i \in \mathcal{V}(u_j)} \mv_{v_i \to u_j}  ||\gv_{u_j}], \thetav_{u})
	\label{eq:node_u_up}
\end{equation}
where $\thetav_{u}$ are the trainable parameters of the \gls{FN} update.

We initialize $\hv_u = \mathbf{0}$ and apply a linear input embedding $\hv_{v_i} = \ell_{\text{ch},i} \times \wv$ where $\wv \in \mathbb{R}^{F_\text{n}}$ is a trainable parameter that projects the scalar $\ell_{\text{ch},i}$ to an $F_\text{n}$-dimensional space. After decoding, the final estimate is given by $\ell_i = \qv^t \hv_{v_i}$, where $\qv \in \mathbb{R}^{F_\text{n}}$ is another trainable vector. Intuitively, projecting messages to higher dimensions can simplify its processing due to the \emph{blessing of dimensionality \cite{donoho2000high}}.

We now perform iterative node and message updates until reaching a maximum of $N_\text{iter}$ iterations. We apply the \emph{flooding} scheduling, i.e., \eqref{eq:msg_vu_up} is applied to all messages in the same time step (the same holds for the other updates in \eqref{eq:msg_uv_up}, \eqref{eq:node_v_up}, \eqref{eq:node_u_up}). Other update schedules are possible, but not further investigated.
Alg.~\ref{alg:gnn_dec} summarizes the detailed decoding algorithm.

\begin{algorithm}[tb]
	\SetAlgoLined
	\SetKwInOut{Input}{Input}
	\SetKwInOut{Output}{Output}
	\SetKwBlock{Repeat}{repeat}{}
	\SetKwFor{RepTimes}{For}{do}{end}
	\SetKwFor{RepTimesshort}{For}{}{}
	\DontPrintSemicolon
	\Input{\glspl{LLR} of the (noisy) channel observations $\ellv_{\text{ch}}$\\
	Set of trainable parameters:\\
	- \Gls{FN} updates $\thetav_{u}$\\
	- \Gls{VN} updates $\thetav_{v}$\\
	- Variable to factor message updates $\thetav_{m_{v \to u}}$\\
	- Factor to variable message updates $\thetav_{m_{u \to v}}$\\
	- \Gls{VN} attributes $\{\gv_{v_i}\} \forall v_i \in \mathcal{V} $\\
	- \Gls{FN} attributes $\{\gv_{u_j}\} \forall u_j \in \mathcal{U} $\\
	- Msg. attr.$\{\gv_{m_{v_i \to u_j}},\gv_{m_{u_j \to v_i}}\},\forall (v_i,u_j) \in \mathcal{E}$\\
	- Input embedding $\wv$\\
	- Output embedding $\qv$
	}
	\Output{Soft estimates $\ellv$ of the codeword bits $\cv$ }

	\% Initialize graph nodes \\
	\RepTimesshort{$v_i \in \mathcal{V}$}{
		$\hv_{v_i} \gets \ell_{\text{ch,i}} \cdot \wv$ \\
	}
	$\hv_u \gets \mathbf{0}$ \\
	\% Run decoding iterations \\
	\RepTimes{$it=0, \dots, N_\text{iter}-1$}{
		\% Update variable to factor node messages \\
		\RepTimesshort{$(v_i,u_j) \in \mathcal{E}$}{
				$\mv_{v_i \to u_j} \gets f^{F_\text{m}}([\hv_{v_i}||\hv_{u_j}||\gv_{v_i \to u_j}], \thetav_{m_{v \to u}})$
		}
		\% Update factor nodes \\
		\RepTimesshort{$u_j \in \mathcal{U}$}{
			$\hv_{u_j} \gets f^{F_\text{n}}([\hv_{u_j}||\frac{1}{|\mathcal{V}(u_j)|}\sum_{v_i \in \mathcal{V}(u_j)} \mv_{v_i \to u_j}  ||\gv_{u_j}], \thetav_{u})$
		}
		\% Update factor to variable node messages \\
		\RepTimesshort{$(v_i,u_j) \in \mathcal{E}$}{
				$\mv_{u_j \to v_i} \gets f^{F_\text{m}}([\hv_{u_j}||\hv_{v_i}||\gv_{u_j \to v_i}], \thetav_{m_{u \to v}})$
		}
		\% Update variable nodes \\
		\RepTimesshort{$v_i \in \mathcal{V}$}{
			$\hv_{v_i} \gets f^{F_\text{n}}([\hv_{v_i}||\frac{1}{|\mathcal{U}(v_i)|}\sum_{u_j \in \mathcal{U}(v_i)} \mv_{u_j \to v_i}  ||\gv_{v_i}], \thetav_{v})$
		}
	}
	\% Project variable node values back to scalar \glspl{LLR} \\
	\RepTimesshort{$v_i \in \mathcal{V}$}{
		$\ell_i \gets \qv^t \hv_{v_i}$
	}
	\Return $\ellv = [\ell_0, \dots, \ell_{n-1}]$ %to check: vector vs. scalar!

	\caption{\gls{GNN}-based Decoding Algorithm}
	\label{alg:gnn_dec}
	\vspace*{0.3cm}
\end{algorithm}

%Remark: please note that the terminology in the ML community is not necessarily aligned the the communications community. We thus aim to keep the terminology as close as possible to the communications community.

Please note that the number of trainable weights is small when compared to state-of-the-art in literature (e.g., the BCH \gls{GNN}-decoder is fully defined by 9640 parameters) and the computational complexity scales linearly in $n$. However, the proposed decoder is still of high complexity as the graph inference requires the execution of the message update functions for every incoming and outgoing edge in the graph. We leave it open for future research to optimize the decoding complexity.

\subsection{Training methodology}

We train the \gls{GNN} by applying iterative loop-unrolling \cite{nachmani2016learning}.
All \gls{NN} weights use the \emph{Glorot uniform} initalizer \cite{glorot2010understanding} while the node/edge attributes are set to $\mathbf{0}$ in our experiments.
The \gls{BCE} loss is implemented as multi-loss over all iterations (e.g., as proposed in \cite{nachmani2016learning})
\begin{equation}
	\begin{split}
\mathcal{L} = \frac{1}{n N_\text{iter}} \sum_{it=0}^{N_\text{iter}-1} \sum_{i=0}^{n-1} & \left[ c_i \operatorname{log}\left(\sigma\left(\ell_i^{it}\right)\right) \right.\\
& \left.+ (1-c_i) \operatorname{log}\left(1-\sigma\left(\ell_i^{it}\right)\right) \right]
\label{eq:bce_loss}
	\end{split}
\end{equation}
where the superscript \emph{it} denotes the decoder output $\ellv$ of the $it$th iteration and $\sigma(x)=\frac{1}{1+e^{-x}}$.
As can be seen later in Fig.~\ref{fig:ber_bch}, this allows for inference of the trained decoder with a variable number of iterations and enables the possibility of early stopping.

We use standard \gls{SGD}-based training with the Adam optimizer.
Please note that the batchsize can be (relatively) small, as the same weights are applied to all nodes, which effectively multiplies the batchsize by $|\mathcal{U}|$ and $|\mathcal{V}|$, respectively (and $|\mathcal{E}|$ for the message updates).
The training requires the availability of an encoder, as all-zero codeword based training is not possible. Thus, we draw new random codewords for every \gls{SGD} iteration.

\begin{table}
	\vspace*{0.06cm}
	\caption{Hyperparameter search for the (63,45) BCH code}\label{tab:hyperparameter}
	\centering
	\begin{tabular}{lll}
		Parameter & Search space & Best\\
		\hline
		\hline
		activation & \{ReLU, sigmoid, tanh\} & tanh\\
		$N_\text{iter}$ & $1-20$ & 8 \\
		$F_\text{n}$ & $1-32$ & 20  \\
		$F_\text{m}$ & $1-32$ & 20  \\
		\# hidden units MLP & $4-128$ & 40\\
		\# MLP layers & $1-4$ & 2 \\
		aggregation function & \{mean, sum\} & mean\\
		\gls{LLR} clipping & $\{10, 16, 20, \infty\}$ & $\infty$ \\
		\hline
		learning rate & - & $10^{-3}-10^{-5}$\\
		batchsize & - & 256\\
	\end{tabular}
\end{table}

As the design space of the hyperparameters is virtually infinite, we have utilized a hyperparameter search \cite{liaw2018tune} and, therefore, evaluated 436 different models. Tab.~\ref{tab:hyperparameter} summarizes the search space and best hyperparameters.
For the simulated BCH code, we did not observe any gains by having edge and node attributes and, thus, set these parameters to $\mathbf{0}$. However, it is instructive to realize that the weights of the node and message update functions are shared between all nodes (and all iterations). Thus, learning edge (or node) specific properties of a specific graph (e.g., learning to remove short cycles) requires the availability of these attributes.

One can make a few interesting observations:
\begin{itemize}
	\item \textbf{Activation:} in particular for small \gls{MLP} sizes, \emph{tanh} often performs better than the \emph{ReLU} and (logistic) \emph{sigmoid} activation.
 	\item \textbf{Wide vs. deep:} we found multiple architectures with a similar \gls{BER} performance, however, some are deeper (i.e., more iterations but less weights) while others are wider (i.e., less iterations required but more weights per update). This gives an additional degree of freedom to the system designer.
  	\item \textbf{Node aggregation:} the \emph{mean} aggregation function tends to show a better performance than the \emph{sum} aggregation and we empirically observed an improved stability when changing the number of iterations \emph{after} training.
  	\item \textbf{Bias:} for scenarios with punctured nodes (e.g., due to rate-matching) deactivating the bias in the MLP layers for node/edge update functions improves the stability for \emph{exact} zero input \gls{LLR} messages as it automatically leads to an all-zero output.
\end{itemize}

\section{Results}

We now evaluate the decoding performance for BCH and \gls{LDPC} codes.
For both codes, no node/edge attributes are used.

\subsection{BCH codes}\label{sec:bch_results}

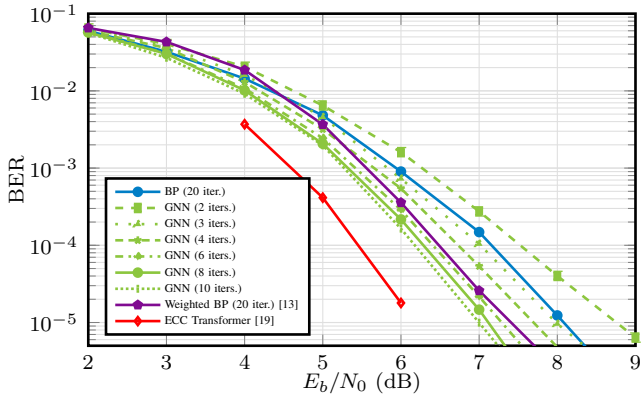
\begin{figure}
	\centering
	\begin{tikzpicture}
	\pgfplotsset{compat=1.5}
	\tikzset{font={\fontsize{8pt}{8}\selectfont}}
	\begin{axis}[
		xmode=normal,
		ymode=log,
		xlabel=$E_b/N_0~(\mathrm{dB})$,
		ylabel=$\mathrm{BER}$,
		y label style={at={(axis description cs:-0.1,.5)},anchor=south},
		x label style={at={(axis description cs:0.5,-0.05)},anchor=north},
		xmin = 2,
		xmax = 9,
		ymax = 1e-1,
		ymin = 5e-6,
		mark size=1.5pt,
		legend style={nodes={scale=0.56, transform shape}},
		legend pos = south west,
		grid=both,
		minor grid style={gray!25},
		major grid style={gray!25},
		width=\columnwidth,
		height=6cm,
		cycle list name=corporate colours markers,
		legend cell align={left},
		line width=1pt]

		%\addplot+ [gray, mark options={fill=gray}]
		%table[x=snr, y=uncoded, col sep=comma]{results/BCH_small.txt};
		%\addlegendentry{Uncoded}

		\addplot+ [mittelblau, mark options={fill=mittelblau}]
		table[x=snr, y=bp-20, col sep=comma]{results/BCH.txt};
		\addlegendentry{BP (20 iter.)}

		\addplot+ [apfelgruen, dashed, mark options={fill=apfelgruen}]
		table[x=snr, y=gnn-2, col sep=comma]{results/BCH.txt};
		\addlegendentry{GNN (2 iters.)}

		\addplot+ [apfelgruen, loosely dotted, mark options={fill=apfelgruen}]
		table[x=snr, y=gnn-3, col sep=comma]{results/BCH.txt};
		\addlegendentry{GNN (3 iters.)}

		\addplot+ [apfelgruen, densely dashed, mark options={fill=apfelgruen}]
		table[x=snr, y=gnn-4, col sep=comma]{results/BCH.txt};
		\addlegendentry{GNN (4 iters.)}

		\addplot+ [apfelgruen, dash dot , mark options={fill=apfelgruen}]
		table[x=snr, y=gnn-6, col sep=comma]{results/BCH.txt};
		\addlegendentry{GNN (6 iters.)}

		\addplot+ [apfelgruen, mark options={fill=apfelgruen}]
		table[x=snr, y=gnn-8, col sep=comma]{results/BCH.txt};
		\addlegendentry{GNN (8 iters.)}

		\addplot+ [apfelgruen, densely dotted, mark options={fill=apfelgruen}]
		table[x=snr, y=gnn-10, col sep=comma]{results/BCH.txt};
		\addlegendentry{GNN (10 iters.)}

		\addplot+ [lila, mark options={fill=lila}]
		table[x=snr, y=wbp-20, col sep=comma]{results/BCH.txt};
		\addlegendentry{Weighted BP (20 iter.) \cite{nachmani2016learning}}

		%copied from paper Nachmani et al.
		% \addplot+ [lila, mark options={fill=lila}]
		% table[x=snr, y=wbpbaseline, col sep=comma]{results/BCH_wbp_ref.txt};
		% \addlegendentry{Weighted BP \cite{nachmani2016learning}}

		\addplot+ [red, mark=diamond, mark options={fill=red}]
		table[x=snr, y=eccbest, col sep=comma]{results/BCH_ECC.txt};
		\addlegendentry{ECC Transformer \cite{choukroun2022error}}

		%to verify that there is no SNR shift
		%\addplot+ [red, densely dotted, mark options={fill=red}]
		%table[x=snr, y=eccbaseline, col sep=comma]{results/BCH_ECC.txt};
		%\addlegendentry{ECC (bp baseline, 50 iters) \cite{choukroun2022error}}

	\end{axis}
\end{tikzpicture}
	\caption{\Gls{BER} performance for decoding of the (63,45) BCH code. The \gls{GNN} was trained for $N_\text{iter}=8$ iterations.}
	\label{fig:ber_bch}
	\vspace*{-0.3cm}
\end{figure}

Fig.~\ref{fig:ber_bch} shows the \gls{BER} performance for the learned (63,45) BCH decoder. The \gls{GNN} hyperparameters are provided in Tab.~\ref{tab:hyperparameter}.
It is worth mentioning that \gls{BP} decoding of BCH codes is known to be highly suboptimal due to the high density of $\Hm$.
Nonetheless, it has become a common benchmark in the literature \cite{nachmani2016learning,choukroun2022error} to decode the (63,45) BCH code, and we therefore provide results for this class of codes. The sub-optimality of BCH provides an interesting playground for our \gls{GNN} experiments as--in contrast to well-designed LDPC codes--\gls{BER} performance gains when compared to \gls{BP} are still possible. One of the currently best \emph{learned} decoders is reported in \cite{choukroun2022error}. However, the number of trainable parameters is 2-3 orders of magnitude larger compared to our decoder.

Although the decoder is trained for a fixed number of decoding iterations $N_\text{iter}=8$, the results in Fig.~\ref{fig:ber_bch} show that the same weights can be used for a variable number of iterations $N_\text{iter}=\{2,3,4,6,8,10\}$ (with a degraded performance for fewer iterations, which is the expected behavior). This allows for the implementation of early stopping at any time. Somewhat to our surprise, the \gls{GNN}-based decoder with only three decoding iterations achieves a similar \gls{BER} performance when compared to the \emph{conventional} BP decoder with 20 iterations.

\subsection{Regular LDPC codes}

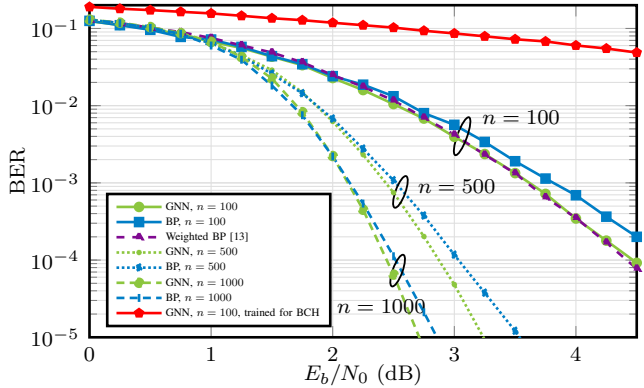
\begin{figure}
	\centering
	\begin{tikzpicture}
	\pgfplotsset{compat=1.5}
	\tikzset{font={\fontsize{8pt}{8}\selectfont}}
	\begin{axis}[
		xmode=normal,
		ymode=log,
		xlabel=$E_b/N_0~(\mathrm{dB})$,
		ylabel=$\mathrm{BER}$,
		y label style={at={(axis description cs:-0.1,.5)},anchor=south},
		x label style={at={(axis description cs:0.5,-0.05)},anchor=north},
		xmin = 0,
		xmax = 4.5,
		ymax = 2e-1,
		ymin = 1e-5,
		mark size=1.5pt,
		legend style={nodes={scale=0.53, transform shape}},
		legend pos = south west,
		grid=both,
		minor grid style={gray!25},
		major grid style={gray!25},
		width=\columnwidth,
		height=6.0cm,
		cycle list name=corporate colours markers,
		legend cell align={left},
		line width=1pt]

		\addplot+ [apfelgruen, mark options={fill=apfelgruen}]
		table[x=snr, y=gnn_n100, col sep=comma]{results/LDPC_reg36.txt};
		\addlegendentry{GNN, $n=100$}

		\addplot+ [mittelblau, mark options={fill=mittelblau}]
		table[x=snr, y=bp_n100, col sep=comma]{results/LDPC_reg36.txt};
		\addlegendentry{BP, $n=100$}

		\addplot+ [lila, dashed, mark options={fill=lila}]
		table[x=snr, y=wbp-20, col sep=comma]{results/LDPC_reg36.txt};
		\addlegendentry{Weighted BP \cite{nachmani2016learning}}

		\addplot+ [apfelgruen, densely dotted, mark options={fill=apfelgruen}]
		table[x=snr, y=gnn_n500, col sep=comma]{results/LDPC_reg36.txt};
		\addlegendentry{GNN, $n=500$}

		% \addplot+ [mittelblau, mark options={fill=mittelblau}]
		% table[x=snr, y=bp-10, col sep=comma]{results/LDPC_reg36_n500.txt};
		% \addlegendentry{BP-10, $n=500$}

		\addplot+ [mittelblau, densely dotted, mark options={fill=mittelblau}]
		table[x=snr, y=bp_n500, col sep=comma]{results/LDPC_reg36.txt};
		\addlegendentry{BP, $n=500$}

		% n = 1000
		% \addplot+ [apfelgruen, mark options={fill=apfelgruen}]
		% table[x=snr, y=gnn-10, col sep=comma]{results/LDPC_reg36_n1000.txt};
		% \addlegendentry{GNN-10, $n=1000$}

		\addplot+ [apfelgruen, densely dashed, mark options={fill=apfelgruen}]
		table[x=snr, y=gnn_n1000, col sep=comma]{results/LDPC_reg36.txt};
		\addlegendentry{GNN, $n=1000$}

		% \addplot+ [mittelblau, mark options={fill=mittelblau}]
		% table[x=snr, y=bp-10, col sep=comma]{results/LDPC_reg36_n1000.txt};
		% \addlegendentry{BP-10, $n=1000$}

		\addplot+ [mittelblau, densely dashed, mark options={fill=mittelblau}]
		table[x=snr, y=bp_n1000, col sep=comma]{results/LDPC_reg36.txt};
		\addlegendentry{BP, $n=1000$}

		% n = 100 weights BCH
		\addplot+ [red, mark options={fill=red}]
		table[x=snr, y=gnn_n100_bch, col sep=comma]{results/LDPC_reg36.txt};
		\addlegendentry{GNN, $n=100$, trained for BCH}

		\node at (axis cs:3.55,0.007) {$n=100$};
		\draw[black,line width=0.7pt,rotate=-20] (axis cs:3.4,0.25) circle (4 and .6);

		\node at (axis cs:3.02,0.0009) {$n=500$};
		\draw[black,line width=0.7pt,rotate=-20] (axis cs:3.08,0.028) circle (4 and .5);

		\node at (axis cs:2.4,0.000025) {$n=1000$};
		\draw[black,line width=0.7pt,rotate=-20] (axis cs:3.28,0.003) circle (4 and .5);

	\end{axis}
\end{tikzpicture}
	\caption{\Gls{BER} performance for decoding of regular (3,6) LDPC codes of the same coderate $r=0.5$. All decoders use $N_\text{iter}=20$ iterations. The \gls{GNN} is only trained once for $n=100$.}
	\label{fig:ber_ldpc_reg}
\end{figure}

To further investigate the scalability, we train the \gls{GNN} to decode regular
$(v=3,c=6)$ \gls{LDPC} codes with \gls{VN} degree $v$ and \gls{CN} degree $c$.
The parity-check matrix is randomly constructed and not further optimized for \gls{BP} decoding. However, the same parity-check matrix is used for both decoders. The results are shown in Fig.~\ref{fig:ber_ldpc_reg}. Note that the \gls{GNN} is trained for a codeword length of $n=100$ and $N_\text{iter}=10$ iterations. For faster training convergence and to avoid another expensive hyperparameter optimization, we increase the \gls{NN} dimensions to approximately 28,700 trainable weights.
As can be seen, the \gls{GNN} decoder slightly outperforms \gls{BP} decoding even for long codes which underlines the scalability of the \gls{GNN}. However, the observed gains tend to diminish with increasing $n$.

To investigate the impact of different graphs, we re-use the trained weights of the BCH decoder from Sec.~\ref{sec:bch_results} to decode the \gls{LDPC} code. For this, we replace the underlying graph structure but keep the weights from the previous training. As can be seen, the learned \gls{GNN} decoder is unfortunately not universal and requires re-training to adopt to the new code.

\subsection{5G NR LDPC codes}

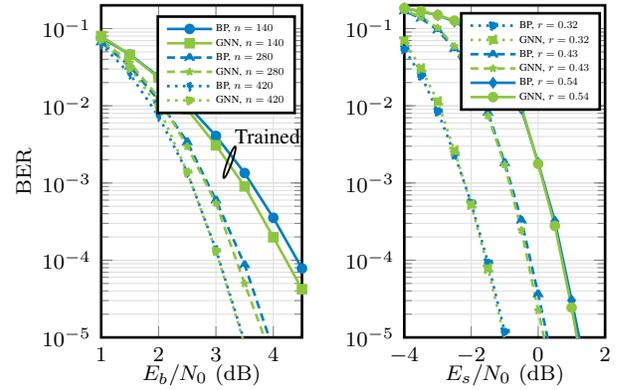
\begin{figure}
	\centering
	\begin{subfigure}[b]{0.48\columnwidth}
		\begin{tikzpicture}
	\pgfplotsset{compat=1.5}
	\tikzset{font={\fontsize{8pt}{8}\selectfont}}
	\begin{axis}[
		xmode=normal,
		ymode=log,
		xlabel=$E_b/N_0~(\mathrm{dB})$,
		ylabel=$\mathrm{BER}$,
		y label style={at={(axis description cs:-0.3,.5)},anchor=south},
		x label style={at={(axis description cs:0.5,-0.05)},anchor=north},
		xmin = 1,
		xmax = 4.5,
		ymax = 2e-1,
		ymin = 1e-5,
		mark size=1.5pt,
		legend style={nodes={scale=0.5, transform shape}},
		legend pos = north east,
		grid=both,
		minor grid style={gray!25},
		major grid style={gray!25},
		width=\columnwidth,
		height=6.0cm,
		cycle list name=corporate colours markers,
		legend cell align={left},
		line width=1pt]

		% k,n=60,140
		% \addplot+ [mittelblau, mark options={fill=mittelblau}]
		% table[x=snr, y=bp-10, col sep=comma]{results/LDPC_5G_k60_n140.txt};
		% \addlegendentry{BP, $n=140$}

		% \addplot+ [apfelgruen, mark options={fill=apfelgruen}]
		% table[x=snr, y=gnn-10, col sep=comma]{results/LDPC_5G_k60_n140.txt};
		% \addlegendentry{GNN,  $n=140$}

		\addplot+ [mittelblau, mark options={fill=mittelblau}]
		table[x=snr, y=bp_k60_n140, col sep=comma]{results/LDPC_5G.txt};
		\addlegendentry{BP, $n=140$}

		\addplot+ [apfelgruen, mark options={fill=apfelgruen}]
		table[x=snr, y=gnn_k60_n140, col sep=comma]{results/LDPC_5G.txt};
		\addlegendentry{GNN,  $n=140$}

		% k,n=120,280
		% \addplot+ [mittelblau, densely dashed, mark options={fill=mittelblau}]
		% table[x=snr, y=bp-10, col sep=comma]{results/LDPC_5G_k120_n280.txt};
		% \addlegendentry{BP, $n=280$}

		% \addplot+ [apfelgruen, densely dashed, mark options={fill=apfelgruen}]
		% table[x=snr, y=gnn-10, col sep=comma]{results/LDPC_5G_k120_n280.txt};
		% \addlegendentry{GNN, $n=280$}

		\addplot+ [mittelblau, densely dashed, mark options={fill=mittelblau}]
		table[x=snr, y=bp_k120_n280, col sep=comma]{results/LDPC_5G.txt};
		\addlegendentry{BP, $n=280$}

		\addplot+ [apfelgruen, densely dashed, mark options={fill=apfelgruen}]
		table[x=snr, y=gnn_k120_n280, col sep=comma]{results/LDPC_5G.txt};
		\addlegendentry{GNN, $n=280$}

		% % k,n=180,420
		% \addplot+ [mittelblau, dotted, mark options={fill=mittelblau}]
		% table[x=snr, y=bp-10, col sep=comma]{results/LDPC_5G_k180_n420.txt};
		% \addlegendentry{BP, $n=420$}

		% \addplot+ [apfelgruen, dotted, mark options={fill=apfelgruen}]
		% table[x=snr, y=gnn-10, col sep=comma]{results/LDPC_5G_k180_n420.txt};
		% \addlegendentry{GNN, $n=420$}

		\addplot+ [mittelblau, dotted, mark options={fill=mittelblau}]
		table[x=snr, y=bp_k180_n420, col sep=comma]{results/LDPC_5G.txt};
		\addlegendentry{BP, $n=420$}

		\addplot+ [apfelgruen, dotted, mark options={fill=apfelgruen}]
		table[x=snr, y=gnn_k180_n420, col sep=comma]{results/LDPC_5G.txt};
		\addlegendentry{GNN, $n=420$}

		\node at (axis cs:3.9,0.004) {Trained};
		\draw[black,line width=0.7pt,rotate=-20] (axis cs:4.3,0.018) circle (4 and .5);

	\end{axis}
\end{tikzpicture}
		\caption{varying length; $r=0.43$}
		\label{fig:ber_ldpc_5g_n}
	\end{subfigure}
	\begin{subfigure}[b]{0.48\columnwidth}
		\begin{tikzpicture}
	\pgfplotsset{compat=1.5}
	\tikzset{font={\fontsize{8pt}{8}\selectfont}}
	\begin{axis}[
		xmode=normal,
		ymode=log,
		xlabel=$E_s/N_0~(\mathrm{dB})$,
		%ylabel=$\mathrm{BER}$,
		y label style={at={(axis description cs:-0.3,.5)},anchor=south},
		x label style={at={(axis description cs:0.5,-0.05)},anchor=north},
		xmin = -4,
		xmax = 2,
		ymax = 2e-1,
		ymin = 1e-5,
		mark size=1.5pt,
		legend style={nodes={scale=0.5, transform shape}},
		grid=both,
		minor grid style={gray!25},
		major grid style={gray!25},
		width=\columnwidth,
		height=6.0cm,
		cycle list name=corporate colours markers,
		legend cell align={left},
		line width=1pt]

		% k,n=120,280
		% \addplot+ [mittelblau, densely dashed, mark options={fill=mittelblau}]
		% table[x=snr, y=bp-10, col sep=comma]{results/LDPC_5G_k120_n280.txt};
		% \addlegendentry{BP, $r=0.43$}

		% \addplot+ [apfelgruen, densely dashed, mark options={fill=apfelgruen}]
		% table[x=snr, y=gnn-10, col sep=comma]{results/LDPC_5G_k120_n280.txt};
		% \addlegendentry{GNN, $r=0.43$}

		\addplot+ [mittelblau, dotted, mark options={fill=mittelblau}]
		table[x=snr, y=bp_k90_n280, col sep=comma]{results/LDPC_5G_EsN0.txt};
		\addlegendentry{BP, $r=0.32$}

		\addplot+ [apfelgruen, dotted, mark options={fill=apfelgruen}]
		table[x=snr, y=gnn_k90_n280, col sep=comma]{results/LDPC_5G_EsN0.txt};
		\addlegendentry{GNN, $r=0.32$}

		\addplot+ [mittelblau, densely dashed, mark options={fill=mittelblau}]
		table[x=snr, y=bp_k120_n280, col sep=comma]{results/LDPC_5G_EsN0.txt};
		\addlegendentry{BP, $r=0.43$}

		\addplot+ [apfelgruen, densely dashed, mark options={fill=apfelgruen}]
		table[x=snr, y=gnn_k120_n280, col sep=comma]{results/LDPC_5G_EsN0.txt};
		\addlegendentry{GNN, $r=0.43$}

		\addplot+ [mittelblau, mark options={fill=mittelblau}]
		table[x=snr, y=bp_k150_n280, col sep=comma]{results/LDPC_5G_EsN0.txt};
		\addlegendentry{BP, $r=0.54$}

		\addplot+ [apfelgruen, mark options={fill=apfelgruen}]
		table[x=snr, y=gnn_k150_n280, col sep=comma]{results/LDPC_5G_EsN0.txt};
		\addlegendentry{GNN, $r=0.54$}

		% % k,n=150,280
		% \addplot+ [mittelblau, densely dashed, mark options={fill=mittelblau}]
		% table[x=snr, y=bp-10, col sep=comma]{results/LDPC_5G_k150_n280.txt};
		% \addlegendentry{BP, $r=0.54$}

		% \addplot+ [apfelgruen, densely dashed, mark options={fill=apfelgruen}]
		% table[x=snr, y=gnn-10, col sep=comma]{results/LDPC_5G_k150_n280.txt};
		% \addlegendentry{GNN, $r=0.54$}

		% % k,n=90,280
		% \addplot+ [mittelblau, densely dashed, mark options={fill=mittelblau}]
		% table[x=snr, y=bp-10, col sep=comma]{results/LDPC_5G_k90_n280.txt};
		% \addlegendentry{BP, $r=0.32$}

		% \addplot+ [apfelgruen, densely dashed, mark options={fill=apfelgruen}]
		% table[x=snr, y=gnn-10, col sep=comma]{results/LDPC_5G_k90_n280.txt};
		% \addlegendentry{GNN, $r=0.32$}

	\end{axis}
\end{tikzpicture}
		\caption{varying coderate; $n=280$}
		\label{fig:ber_ldpc_5g_r}
	\end{subfigure}
	\caption{\Gls{BER} performance for decoding of the 5G NR LDPC code \cite{38212} for $N_\text{iter}=10$ decoding iterations. The \gls{GNN} is only trained once for $(k,n)=(60,140)$.}
	\label{fig:ber_ldpc_5g}

\end{figure}

We now train the \gls{GNN} for decoding of the 5G NR \gls{LDPC} code \cite{38212} and change the length \emph{after} training by changing the lifting factor of the code. The \gls{GNN} consists of 18,900 trainable weights. The results are shown in Fig.~\ref{fig:ber_ldpc_5g}. %We want to emphasize that the \emph{macro} parameters of the code remain the same (same degree profile, same coderate).
As the 5G NR \gls{LDPC} code inherently requires rate-matching, we train the \gls{GNN} on the underlying (non-rate-matched) parity-check matrix and apply rate-matching only for inference. The rate-matching itself is not learned.
This avoids that the first $2Z$ information bits are punctured during the training (where $Z$ denotes the lifting factor of the code). Otherwise, we empirically observed instabilities during training \emph{through} the rate-matching. The results in Fig.~\ref{fig:ber_ldpc_5g} show that both decoders achieve the same decoding performance for a wide range of code parameters and for the exact code parameters used for training even a small gain can be observed. Keep in mind that the 5G \gls{LDPC} code is specifically optimized for BP decoding and the evaluation is done for the \gls{AWGN} channel. Intuitively, there is only little room for improvement through deep learning for this optimal setting.

As shown in Fig.~\ref{fig:ber_ldpc_5g_n}, the same weights can be scaled to longer codeword lengths and the same flexibility can be also observed w.r.t. the coderate in Fig.~\ref{fig:ber_ldpc_5g_r}. This could enable a simplified training based on the underlying protograph of the LDPC code and only a few sets of different weights must be stored to support all possible code parameters of a given standard. However, we would like to emphasize that the training--in particular, when rate-matching is involved-- requires carefully adjusted hyperparameters which seem to be highly code specific.
Note that both algorithms use \emph{flooding} scheduling for their node updates and the required number of iterations may change for a different scheduling.

\section{Discussion and open issues}

As mentioned above, this work raises more questions than it answers and, thus, we would like to state a few of them for possible future research:
\begin{itemize}
	\item Is training of a universal GNN decoder possible? One possible way to achieve such a universality is the training of the same \gls{GNN} decoder weights for multiple \gls{FEC} codes in parallel.
	\item The effect of edge/node attributes is currently unknown. We empirically observed that the decoder does not (strongly) benefit from these attributes, but the decoding and memory complexity increases. However, when compensating for specific weaknesses of the decoding graph, these attributes may be beneficial.
	\item The decoding complexity of the proposed \gls{GNN} is higher when compared to \gls{BP} as the \glspl{MLP} must be executed many times. Could different architectures for the update functions reduce the complexity?
	\item This work could be extended towards heterogenous \glspl{GNN} such that for irregular node degrees different trainable functions are applied.
	\item Apply graph modifications and pruning. We empirically observe that the decoding graph is still crucial for the decoding performance. Similar to \cite{buchberger2020pruning}, pruning of the decoding graph could be applied during training.
	\item Extensions to non-binary codes and non-AWGN channels are possible, and we expect higher gains compared to \gls{BP} decoding when operating on real-world datasets including different kinds of impairments.
\end{itemize}

\section{Conclusion}
We have presented a \gls{GNN}-based algorithm for decoding of binary linear block codes where the code structure can be directly embedded into the decoding algorithm. As a result, we achieved a competitive bit-error-rate performance for BCH codes with significantly fewer decoding iterations when compared to \emph{classical} \gls{BP} decoding. For \gls{LDPC} codes, the performance remains competitive with \gls{BP} decoding for the \gls{AWGN} channel, however, adapting the decoder to different channels or even to real-world datasets is straightforward. In order not to over-sell our results, we would like to emphasize that the computational complexity of the proposed decoder is high. However, for scenarios where no suitable \emph{classical} decoder is available (such as for quantum error control coding \cite{liu2019neural}), it opens an interesting avenue for future research.
It remains also open how to reduce the overall decoding complexity and, thus, to improve the energy-efficiency. It seems promising to combine the presented decoder with optimized~\gls{NN} libraries for adaptive re-training of small networks, such as \cite{tiny-cuda-nn}.

\bibliographystyle{IEEEtran}
\bibliography{IEEEabrv,bibliography}

\end{document}